\documentclass[12pt]{article}
\usepackage{amsmath,latexsym}
\usepackage{graphicx}
\begin{document}

 \title{\Huge Galactic rotation curves and strange quark matter with observational constraints }
 \author{M. Kalam$^1$,F.Rahaman$^2$,Sk. M. Hossein$^3$ and J.Naskar$^4$  }
\date{}
 \maketitle

 \begin{abstract}
 We obtain the space time of the galactic core in the framework of
general relativity by taking the flat rotational curve as input
and considering the matter content in the galactic core region as
strange quark matter. We also obtain the energy density, radial
and tangential pressures of the core strange quark matter.
Significantly we have shown that Bag Constant takes an important
role to stabilize the circular orbit of the test particles. We
also give a limit of the Bag Constant for the existence of quark
matter in the galactic halo region.
\end{abstract}

  \footnotetext{


$^1$Department of Physics, Aliah University, Sector-V, Salt Lake, Kolkata- 700091,India.\\
E-Mail:mehedikalam@yahoo.co.in ; kalam@iucaa.ernet.in

$^2$Dept.of Mathematics, Jadavpur University, Kolkata-700 032, India.\\
 E-Mail:farook\_rahaman@yahoo.com ; rahaman@iucaa.ernet.in

 $^3$
Department of Mathematics, Aliah University, Sector-V, Salt Lake, Kolkata- 700091,India.\\
E-Mail:sami\_milu@yahoo.co.uk

$^4$Department of Physics, Aliah University, Sector-V, Salt Lake, Kolkata,India.\\
E-Mail:jayanta86@gmail.com

    \mbox{} \hspace{.2in}}

\section{Introduction}
Presence of dark matter was suspected  first by
Oort\cite{Oort30a, Oort30b, Oort30c}. Later Zwicky \cite{fZ33,
fZ37}  concluded that a large amount of non-luminous matter
remains hidden in the galactic haloes. Observation of flatness of
galactic rotation curves \cite{kF70, RR73, OPY74, EKS74, RTF78,
SR01}  was consequitively confirmed the  Zwicky's suggestion
regarding the presence of dark matter. Afterwards,
gravitational lensing of objects like bullet clusters and the
temperature distribution of hot gas in galaxies and galactic
clusters have further confirmed the existence of dark matter.The
flatness of the galactic rotation curves  indicates  the presence
of much more matter within the galaxy than the visible
material.Estimation supports a spherical distribution of matter
(galaxy) is surrounded by a dark matter halo.
Gravitational effect of dark matter is more manifest at larger
radius\cite{Bergeman02}. Now a days, the amount of dark matter present in the
Universe has become known in more precise manner. Cosmic
Microwave Background Radiation  data indicates that
nearly $85\%$  of total matter in the galaxy is dark in nature.
Big Bang analysis and cosmological observational data indicates
that the bulk of dark matter may be cold or warm,stable or
long-lived and non-interacting with the visible matter.However,
after a long and through investigations, a little
is yet known about the nature of dark matter.\\

Gell-Mann\cite{GellMann64}and Zweig \cite{Zweig64} had
independently proposed that hadrons are consists of even more
fundamental particles called quarks,a proposition which got
experimental support later on. As quarks are not  free particles,
the quark confinement mechanism have been dealt with great details
in Quantum Chromo Dynamics(QCD). In the MIT bag
model\cite{Chodos74},it was suggested that the quark confinement
is due to a universal pressure $B_g$, which is called the bag
constant. Farhi and Jafee \cite{Farhi84} and Alcock et al
\cite{Alcock86} had shown that the value of the bag constant, $B_g
$ should lie between 60-100
$MeV /fm^{3}$, for a strange quark matter.\\

There are a number of proposed candidates for dark matter. One of
them is standard cold dark matter(SCDM)\cite{ESM90,aP04}.In this
literature, we suggested that quark matter may be such candidate
which is also proposed by several
researchers\cite{Garcia10,Forbes10,Weissenborn11}. Also, it is
well accepted that  quark matter is exist at the centre of neutron
stars, strange
stars\cite{Garcia10,Drake02,Jaikumar06,Rodrigues11,Bordbar11,Madsen99,Weber05}.
Immediately after the Big Bang the Universe underwent a Quark
-Qluon-Plasma(QGP) phase. In the famous Large Hadronic Collider
(LHC) experiment, Scientists recreate the conditions similar to
those encountered before and in the early hadronization
period\cite{EMU01}.As the expansion takes place, the Universe
cools down and the hot Quark-Gluon-Plasma(QGP) freezes slowly to
produce individual hadrons\cite{Fromerth05}.Most acceptable theory
of strong interaction, Quantum Chromo Dynamics (QCD),suggests
that under extreme condition a hadronic system can undergoes a
phase transition from confined hadronic matter to the QGP phase.
It is suggested that the core of neutron stars may consists of
cold QGP\cite{Walker91}.

The purpose of the present work is to show that, with the input
of flat rotation curve and assuming the dark matter contents as
strange quark matter, the gravity in the halo region is
attractive under certain condition where Bag Constant takes an
important role.We  also want to investigate the presence of  quark
matter  in the galactic halo region. This is the main motto of our
work.

\section{Field equations and general results}

 The general static spherically symmetric spacetime is represented  as
\begin{equation}
ds^2 = -e^{\nu(r)} dt^2 + e^{\lambda(r)} dr^2 + r^2 (d \theta ^2 +
sin^2 \theta d\phi^2)
\end{equation}
where $\nu(r)$ and $\lambda(r)$ are the metric potentials and are
function of the space coordinate $r$ only. We further assume that the
energy-momentum tensor for the strange quark matter filling the
interior of the galaxy may be expressed in the standard form as
\[T_{ij}=diag(\rho,-p_r,-p_t,-p_t)\]
where $\rho$ ,$p_r$ and $p_t$ correspond to the energy density,
radial pressure and transverse pressure of the baryonic matter,
respectively.

The Einstein's field equations for the metric (1) in presence of
strange quark matter are then obtained as (with $G=c=1$ under
geometrized relativistic units)
\begin{equation}
e^{-\lambda} \left( \frac{\lambda^{\prime}}{r} - \frac{1}{r^{2}}
\right) + \frac{1}{r^{2}} = 8\pi \rho,
\end{equation}
\begin{equation}
e^{-\lambda} \left( \frac{\nu^{\prime}}{r} + \frac{1}{r^{2}}
\right) - \frac{1}{r^{2}} = 8\pi  p_r,
\end{equation}
\begin{equation}
\frac{1}{2}e^{-\lambda} \left (\nu^{{\prime}{\prime}} +
\frac{{\nu^{\prime}}^{2}}{2} - \frac{{\nu^{\prime}
\lambda^{\prime}}}{2} + \frac{\nu^{\prime} - \lambda^{\prime}}
{r}\right)= 8\pi p_t
\end{equation}

Following the MIT bag model, we express the strange quark matter
equation of state(EOS) in the form
\begin{equation}
p_r=\frac{1}{3}(\rho-4B_g),
\end{equation}
where $B_g$ is the Bag constant(in $MeV/fm^3$ units).

\section{Solutions for the strange quark matter model}
Assuming the known flat rotation curve condition
$v_\phi=[\frac{r(e^\nu)'}{2e^\nu}]^{1/2}=$ constant tangential
velocity\cite{Chandra83,Rahaman07} obtained a solution of it as
\cite{Rahaman08}
\begin{equation}
e^{\nu}= B_0 r^l,
\end{equation}
where $l=2v_\phi^2$ and $B_0$ is an integration
constant.

It is to be noted that the observed rotational velocity
$v^{\phi}$ becomes more or less a constant with $v^{\phi} \sim
10^{-3} $(300 km/s) for a typical galaxy \cite{Matos00,Salucci97}.

Now, from equations (2)-(5) and then using the simple expression in the equation(6) we get the simplified form as
\begin{equation}
-(e^{-\lambda}\lambda') + (4+3l)\frac{e^{-\lambda}}{r}=\frac{4}{r}-32\pi B_g r
\end{equation}

Substituting $e^{-\lambda}=x$, we get
\begin{equation}
x' + \frac{ax}{r}=\frac{4}{r}+cr
\end{equation}
where $a= 4+3l$,$c=-32\pi B_g$. Solving the above equation we get
the metric potential as
\begin{equation}
e^{-\lambda}  =\frac{4a+car^2+8}{a(a+2)}+\frac{D}{r^a}
\end{equation}
where D is integration constant.

Therefore, with the flat rotational curve condition, the metric (1) becomes
\begin{equation}
ds^2 = -B_0r^l dt^2 + \left[\frac{4a+car^2+8}{a(a+2)}+\frac{D}{r^a}\right]^{-1} dr^2 + r^2 (d \theta ^2 +sin^2 \theta d\phi^2)
\end{equation}

Using equation (6) and (9) in (2)-(5), we get
\begin{eqnarray}
 \rho = \frac{1}{8\pi}\left(\frac{D(a-1)}{r^{a+2}}-\frac{3c}{a+2}+\frac{a-4}{ar^2}\right)\\
p_r= \frac{1}{8\pi r^2}\left[\left(\frac{4a+8+car^2}{a(a+2)}+\frac{D}{r^a}\right)\left(l+1\right)-1\right]\\
p_t=\frac{1}{8\pi}\left[\left(\frac{l}{2r}\right)^2\left(\frac{4a+8+car^2}{a(a+2)}+\frac{D}{r^a} \right)+\left(\frac{l}{4}+\frac{1}{2}\right)\left(\frac{2c}{a+2}-\frac{aD}{r^{a+2}}\right)  \right]
\end{eqnarray}

The pressure anisotropy is a good feature from the point of view
of exterior matching. The solution can be matched to the
Schwarzschild exterior metric at the boundary of the
halo\cite{Bharadwaj03}. As the strange matters are not exotic in
nature,  they must satisfies the Null Energy Condition(NEC)
\cite{Hochberg98}. We find that the NEC is satisfies at everywhere
i.e.  $ \rho > 0$ and $ \rho + p_r >0$.  However, we note that
here the  Strong Energy Condition(SEC) is violated ( see figure
1). Therefore, we comment the quark matter   in the galactic halo
region is marginally real like.

\begin{figure*}[thbp]
\begin{tabular}{rl}
\includegraphics[width=5.5cm]{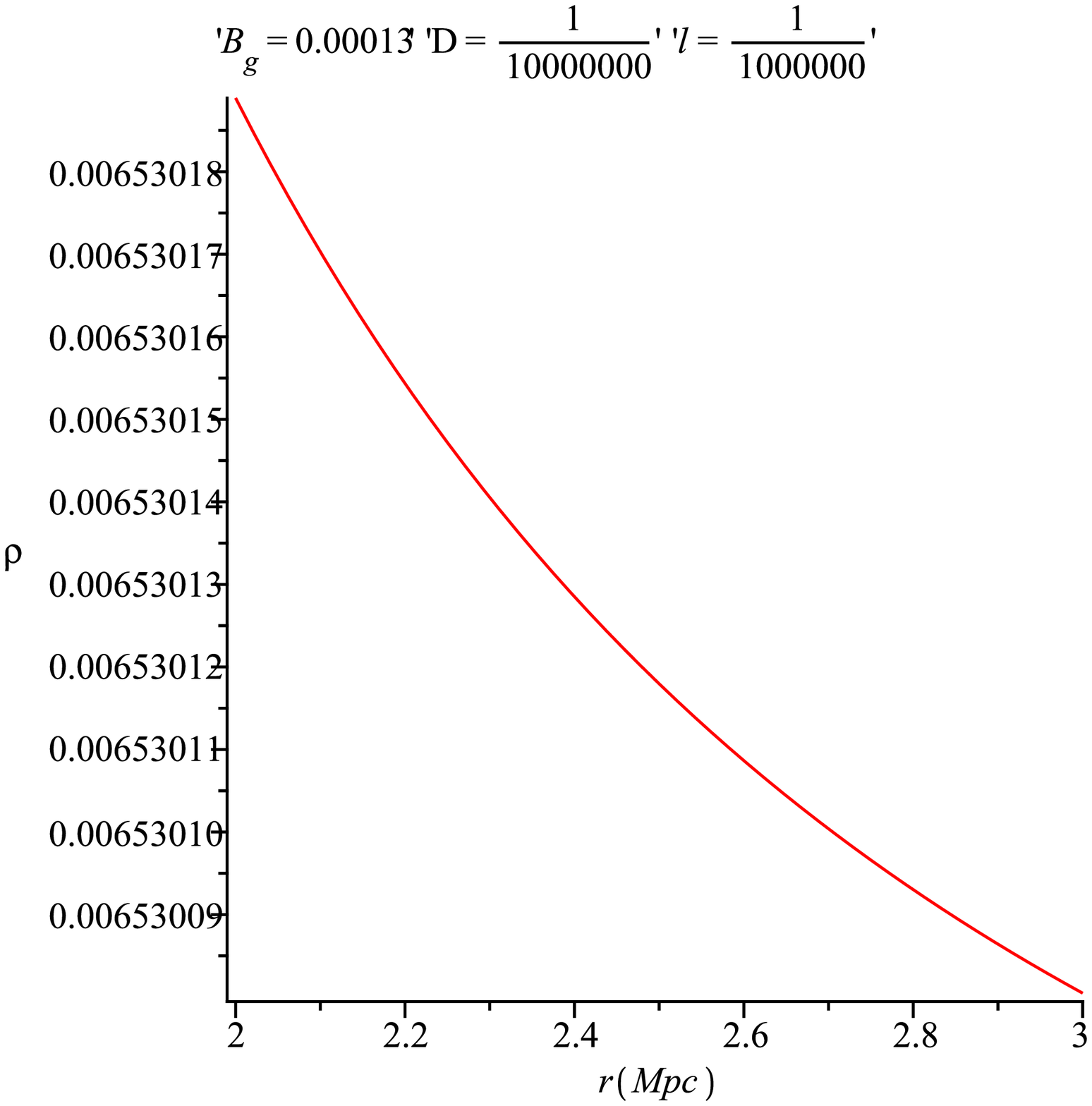}&
\includegraphics[width=5.5cm]{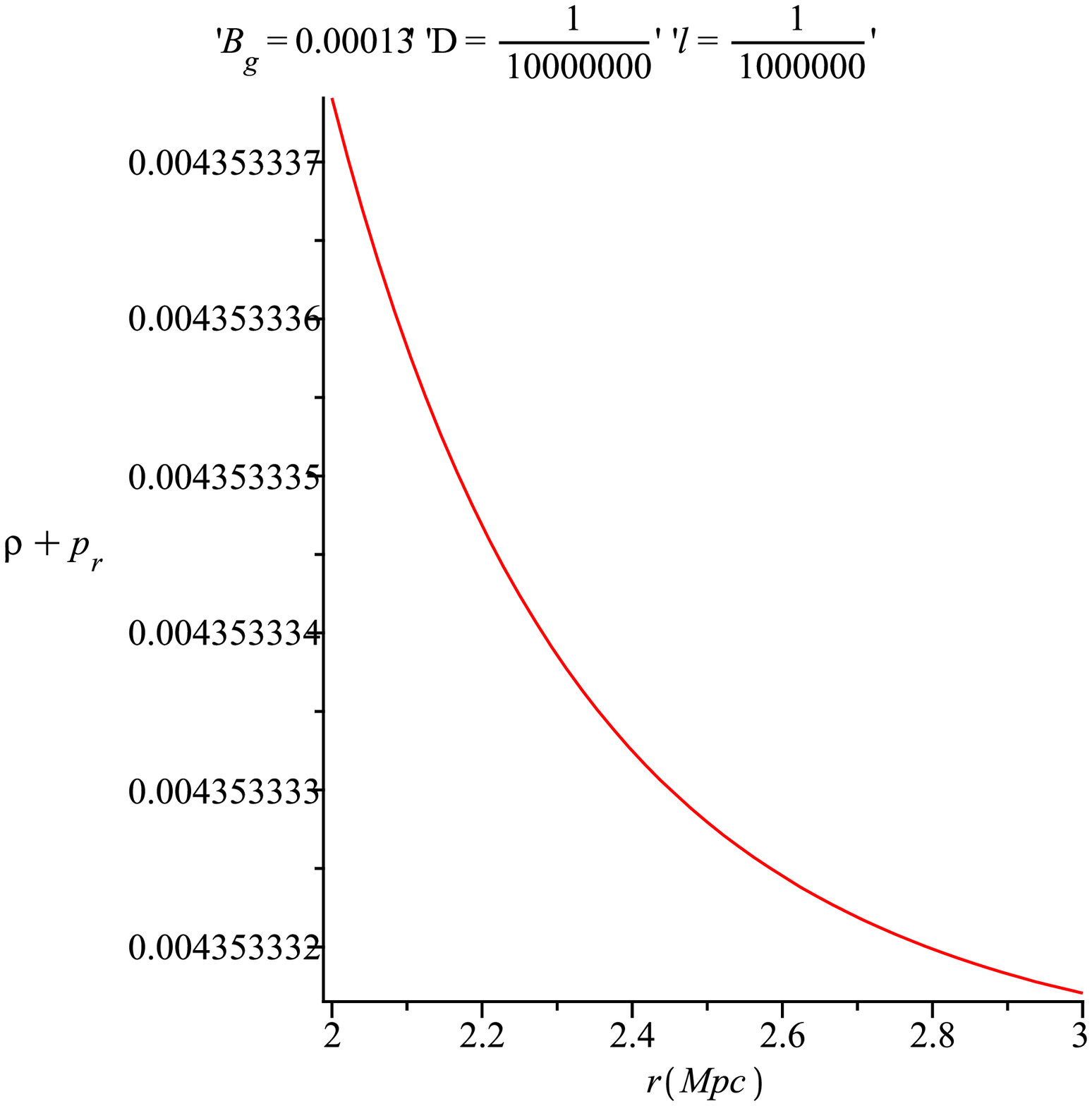}
\includegraphics[width=5.5cm]{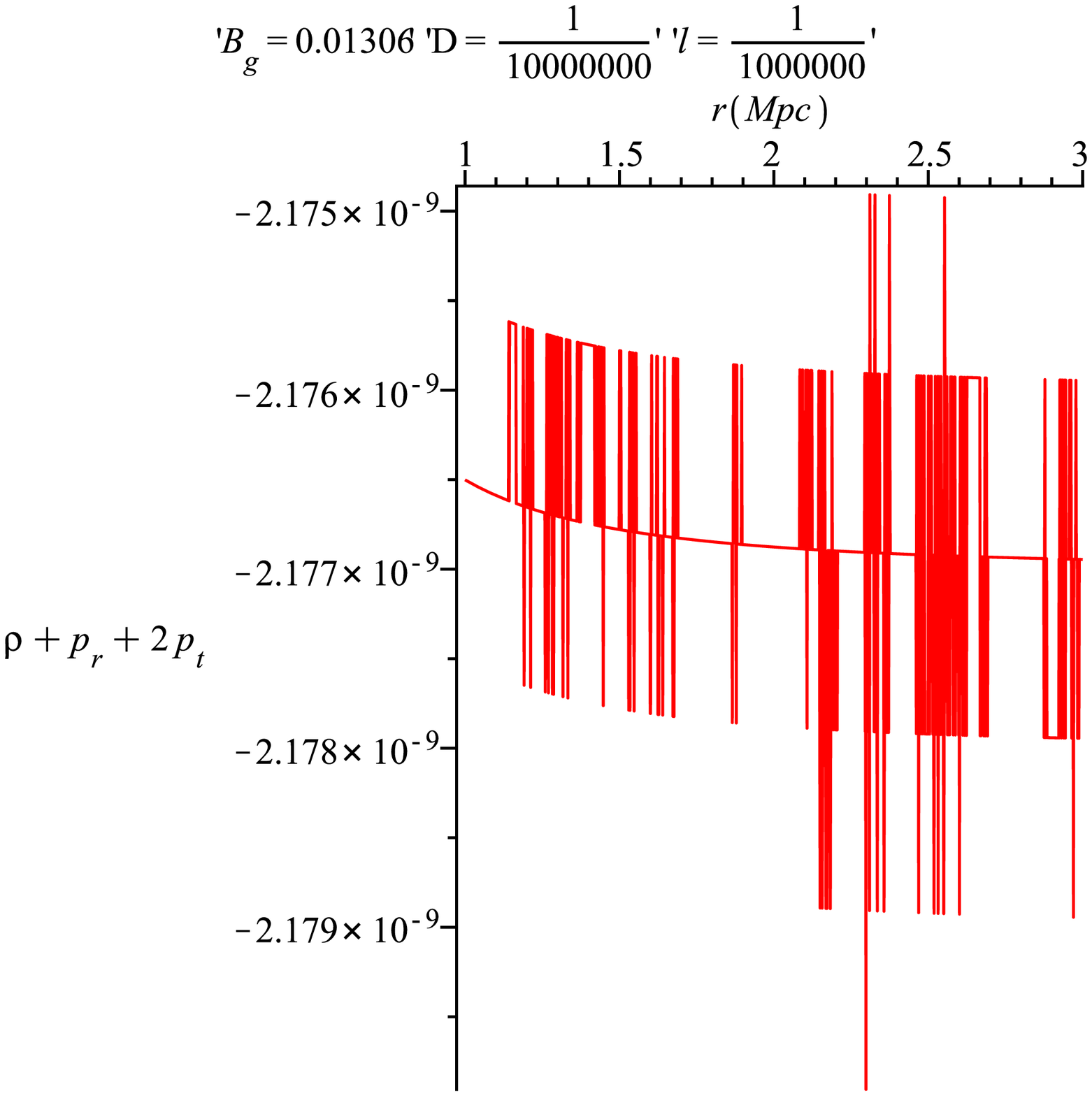} \\
\end{tabular}
\caption{ (\textit{Left})  Variation of $\rho $ with distance
r(Mpc) for the specific values of the parameters, $ D=10^{-7}~,~l=
10^{-6}~,~B_g=.00013$. (\textit{Middle})Variation of $\rho+p_r$
with distance r(Mpc) for the same specific values of the
parameters. (\textit{Right})Variation of $\rho+p_r+2p_t$ with
distance r(Mpc) for the same specific values of the parameters.}
\end{figure*}

\section{Few Features}
\subsection{Stability }

Let, the four velocity, $U^\alpha = \frac{dx^\sigma}{d\tau}$
for a test particle moving solely in the galactic halo region
(restricting ourself to $\theta=\pi/2$), the equation
$g_{\nu\sigma}U^\nu U^\sigma = -m_0^2$ can be cast in a Newtonian
form as\cite{Nandi09}

\begin{equation}
 \left(\frac{dr}{d\tau}\right)^2 = E^2 + V(r)
\end{equation}
which gives
\begin{equation}
V(r) = -\left[E^2\left(1-\frac{r^{-l}[\frac{4a+8+car^2}{a(a+2)}+\frac{D}{r^a} ]}{B_0}   \right)+ \frac{4a+8+car^2}{a(a+2)}+\frac{D}{r^a}\left(1+\frac{L^2}{r^2} \right)\right]\\
\end{equation}
\begin{equation}
E=\frac{U_0}{m_0},  L=\frac{U_3}{m_0}
\end{equation}
where the constants E and L, respectively, represents the conserved relativistic energy and angular momentum per unit mass of the test particle.
Circular orbits of the test particles are defined by $ r= R =$constant, so that $\frac{dR}{d\tau}=0$ and $\frac{dV}{dr}|_{r=R} = 0.$
From these two conditions we get the conserved parameters as: \\
\begin{equation}
L=\pm \sqrt{\frac{l}{2-l}} R
\end{equation}
and using this  in $V(R)= -E^2$, one can easily get,
\begin{equation}
E=\pm \sqrt{\frac{2B_0}{2-l}} R^{\frac{l}{2}}
\end{equation}

For  stable orbit $\frac{d^2V}{dr^2}|_{r=R} < 0$ and for unstable
orbit $\frac{d^2V}{dr^2}|_{r=R} > 0$. Putting the expressions for
L and E in $\frac{d^2V}{dr^2}|_{r=R} $, we get the final
results as

\begin{equation}
\frac{d^2V}{dr^2}|_{r=R}=- \left[\frac{2l}{R^2}\left(\frac{4a+8+caR^2}{a(a+2)}+\frac{D}{R^a}
    \right) + \frac{2(2l-1)}{R(2-l)}\left(\frac{aD}{R^{a+1}}-\frac{2cR}{a+2}  \right)\right]
\end{equation}
Thus $\frac{d^2V}{dr^2}|_{r=R} < 0$ so that circular orbits are
stable under consideration when $D > 0 $ and $B_g<
\frac{(a+2)(2-l)}{64 \pi (2-2l-l^2)} \left[\frac{8l}{aR^2} +
\frac{2D}{R^{a+2}}\left(l-\frac{a(1-2l)}{2-l}\right)\right]$. The
first condition is always satisfied since, from equation(10), we
see that  D has the same dimension as radius, R. Again, as the Bag
Const., $B_g$ and radius R are always positive , so   D is
obviously positive. As this results is in agreement with the
observations, therefore we can conclude that the value of the Bag
constant,$B_g$ should be less than $\frac{(a+2)(2-l)}{64 \pi
(2-2l-l^2)} \left[\frac{8l}{a R^2} +
\frac{2D}{R^{a+2}}\left(l-\frac{a(1-2l)}{2-l}\right)\right]$, if
indeed,  there exists any quark matter present in the galactic
haloes. Therefore from this model, we can give the limit  and a
rough estimation of  the Bag constant within the galactic halo
region. Thus we note that  Bag Const.,$B_g$ plays an important
role to stable the circular orbit.

\subsection{Attraction in Strange quark matter}
Observational indication is  that the gravity on the galactic
scale is attractive. Again, the stable circular orbit indicates
that the particles are being accelerated towards the Centre of the
Galaxy. We can see it by studying the geodesic for a test
particle that has been placed at a circular path of radius $r$.\\
Now,
\begin{equation}
           \frac{d^2x^\alpha}{d\tau^2}
    +\Gamma^{\mu\gamma}_{\alpha}\frac{dx^\mu}{d\tau}
   \frac{dx^\gamma}{d\tau}  = 0.
 \end{equation}
This equation implies that

           $\frac{d^2 r} {d\tau^2} = - \frac{1}{2} \left[\frac{4a+8+car^2}{a(a+2)}
        +\frac{D}{r^a}\right]\left[\left(\frac{Da}{r^{a+1}}-\frac{2car}{a(a+2)}\right)\left(\frac{4a+8+car^2}{a(a+2)}
        +\frac{D}{r^a}\right)^{-2}
         \left(\frac{dr}{d\tau}\right)^2
           + B_0 l r^{l-1}
           \left(\frac{dt}{d\tau}\right)^2\right].$
\begin{equation} \end{equation}

Obviously, the  quantities in the square brackets are positive.
Thus particles are attracted towards the centre with some
boundaries.

\subsection{Gravitational Energy}

We can easily determine the total gravitational energy $E_g$ between
two fixed radii, say, $r_1$ and $r_2$ \cite{Lynden07}:
\begin{multline}
E_{g}=M-E_{M}=4\pi\int_{r_{1}}^{r_{2}}
  [1-\sqrt{e^{\lambda(r)}}]\rho r^{2}dr\\
= 4\pi\int_{r_{1}}^{r_{2}}\left[1-\sqrt{\frac{1}{\frac{4a+car^2+8}{a(a+2)}+\frac{D}{r^a}}}%
\right]\left[\frac{1}{8\pi}\left(\frac{D(a-1)}{r^{a+2}}-\frac{3c}{a+2}+\frac{a-4}{ar^2}\right)\right]r^{2}dr.
\end{multline}
where
\begin{equation}
M=4\pi\int_{r_{1}}^{r_{2}}\rho r^{2}dr
\end{equation}
is the Newtonian mass given by

\begin{equation}
M = 4 \pi \int_{r_{1}}^{r_{2}} \rho r^2 dr =\frac{1}{2}\left
[\frac{Da}{(2-a)r^{a-2}}-\frac{D}{(1-a)r^{a-1}} + \frac{a-4}{a}r-
\frac{cr^3}{a+2}\right]_{r_{1}}^{r_{2}}
\end{equation}

Here, $E_M$ is the sum of the other forms of energy like the rest
energy, kinetic energy, and internal energy.

 For the specific values of the parameters $r_1 =1~, ~r_2=3~,~D=10^{-7}~,~l= 10^{-6}~,~B_g=.00013$, we calculate
 the numerical value of the integrand
(22) describing the total gravitational energy as $ E_g=
-0.00168098$, which indicates that $E_g < 0$, in other words,
there is an
 attractive effect in the halo region.  This result is very much
expected for the matter source in the galactic halo region that
produces stable the circular orbit.

\subsection{Observational Constraints}
We can rewrite the metric(1) in the form
\begin{equation}
ds^2 = -e^{2\Phi(r)} dt^2 + \frac{1}{1-\frac{2m(r)}{r}}dr^2 + r^2
(d \theta ^2 + sin^2 \theta d\phi^2)
\end{equation}
where, the metric functions are given by
\begin{eqnarray}
\Phi(r) = \frac{1}{2}\left(\ln B_{0}+l \ln r\right)\\
m(r) =
\frac{r}{2}\left[1-\frac{4a+car^2+8}{a(a+2)}-\frac{D}{r^a}\right]
\end{eqnarray}
These two functions are not necessarily the same as the potential and
mass functions are found  from the observations.  The
first post-Newtonian approximation \cite{Misner73}, the
gravitational potential $\Phi (r)$,  in general relativity, is
given by
\begin{equation}
\nabla^2 \Phi \approx R_{tt} \approx 4\pi (\rho+p_{r}+2p_{t}).
\end{equation}
 which reduces to the equation
\begin{equation}
\nabla^2\Phi_N = 4\pi \rho
\end{equation}
provided the contributions of the pressures  are negligible in
comparison to energy density.Here, $\Phi_{N}$ indicates the Newtonian potential.
This general relativistic $\Phi$ also appears in the
wavelength shifts $z_{\pm}$  of an emission line of a massive
particle for an edge-on galaxy is \cite{Nucamendi01,Lake04} is given by
\begin{equation}
1+z_{\pm} =
\frac{1}{\sqrt{1-r\Phi'(r)}}\left[\frac{1}{e^{\Phi(r)}}-\frac{\pm
\mid b \mid \sqrt{r\Phi'(r)} }{r}\right]
\end{equation}
where  $\pm$ signs indicates the approaching and receding
particles and b is the impact parameter. The above equation(30)
can approximately written as \cite{Faber06} $ z_{\pm}^2 \approx
r\Phi'(r).$ \\
Therefore, the usual methods for obtaining potential for rotation
curve measurements yields the pseudo-potential
\begin{equation}
\Phi_{RC} = \Phi \neq \Phi_{N}.
\end{equation}
From equation(28) one can say that pseudo-mass,
\begin{equation}
m_{RC} = r^2\Phi'(r)\approx 4\pi\int(\rho +p_{r}+2p_{t})r^2dr =
\frac{lr}{2}
\end{equation}
The pseudo-mass $m_{RC}$ becomes  Newtonian mass M(r)
when pressure contributions are negligible in comparison to energy
density.\\

We can also treat Photons  as probe particles as they interacts
with the gravitational field in the core of the galaxy during
their travel to the observer.The gravitational effect on photon
motion can be measured in terms of a refractive index
$\mu(r)$\cite{Boonserm05}.The geodesic of both massive and
photonlike particles could be expressed exactly in terms of a
single generalized refractive index $ N=\frac{\mu^2v}{c}$ where
$v$ is the three-velocity of the particle. For light motion, $v$ =
$\frac{c}{\mu}$.\\
For unspecified metric functions $\Phi(r)$ and
m(r),\cite{Faber06} argue that $\mu(r) =
1-2\Phi_{lens}+O\left[\Phi^2_{lens}\right]$ where they defined
the lensing pseudo-potential as
\begin{equation}\Phi_{lens} = \frac{1}{2}\Phi(r) +\frac{1}{2}
\int{\frac{m(r)}{r^2}dr}.
\end{equation}
Another pseudo-mass obtained from lensing measurement has been
written as \cite{Faber06}
\begin{equation}
m_{lens} = \frac{1}{2}r^2\Phi'(r)+\frac{1}{2}m(r).
\end{equation}
The first order approximation of Einstein's equation yield
\begin{eqnarray}
\rho(r) \approx \frac{1}{4\pi r^2}[2m'_{lens}(r)-m'_{RC}(r)]\\
4\pi r^2(p_{r}+2p_{t}) \approx 2[m'_{RC}(r)-m'_{lens}(r)]
\end{eqnarray}
where the right-hand sides represents, respectively,
pseudo-density and pseudo-pressures. Furthermore, \cite{Faber06}
defined a dimensionless quantity
\begin{equation}
\omega(r) = \frac{p_{r}+2p_{t}}{3\rho} \approx
\frac{2(m'_{RC}-m'_{lens})}{3(2m'_{lens}-m'_{RC})}.
\end{equation}
where the pseudo-quantities on the right hand side of the
equations (35-37) determine, respectively, the observed
density, pressure and equation of state.\\

It is to be mentioned that the observable quantities are the
pseudo-quantities. The general procedure to determine the metric
are as follows: Once anyone is able to observationally determine
the profiles of pseudo-quantities, one can work backwards to find
the corresponding metric functions. This kind of reverse
technique observational astrophysicists use. If the observed
pseudo-profiles fit (up to experimental error) with the analytic
profiles of a priori given metric functions, one can say that the
solution is physically sustainable. Otherwise, it has to be ruled
out as non-viable. In this sense, the observed pseudo-profiles
play the role of constrains on the possible metric solutions.\\
For the presence situations, we obtain the following constraint
equations on $\Phi(r), m(r)$, the pressure profile and the
equation of state:
\begin{equation}\Phi_{RC} = \frac{1}{2}\left(\ln B_{0}+l \ln r\right)\end{equation}
\begin{equation}m_{RC} = r^2\Phi'(r)= \frac{l r}{2}\end{equation}
\begin{equation}\Phi_{lens} = (l+1-\frac{4}{a})\frac{\ln r}{4}+\frac{1}{4}\ln
B_{0}+\frac{r^2}{4a}\left[\frac{D}{r^{a+2}}-\frac{a
c}{2(a+2)}\right]\end{equation}
\begin{equation}m_{lens} =
\frac{l^2}{8}+\frac{r}{4}\left(1-\frac{4a+8+car^2}{a(a+2)}-\frac{D}{r^a}\right)\end{equation}
\begin{equation}2(m'_{RC}-m'_{lens}) =
2\left[\frac{l}{2}-\frac{1}{4}\left(1-\frac{4a+8+car^2}{a(a+2)}-\frac{D}{r^a}\right)-\frac{r}{4}\left(-\frac{2car}{a(a+2)}+\frac{Da}{r^{a+1}}\right)\right]\end{equation}
\begin{equation}\omega(r) =\frac{2(m'_{RC}-m'_{lens})}{3(2m'_{lens}-m'_{RC})}
=\frac{2\left[\frac{l}{2}-\frac{1}{4}\left(1-\frac{4a+8+car^2}{a(a+2)}-\frac{D}{r^a}\right)
-\frac{r}{4}\left(-\frac{2car}{a(a+2)}+\frac{D
a}{r^{a+1}}\right)\right]}
{6\left[\frac{1}{4}\left(1-\frac{4a+8+car^2}{a(a+2)}-\frac{D}{r^a}\right)+\frac{r}{4}\left(-\frac{2car}{a(a+2)}+\frac{Da}{r^{a+1}}\right)\right]}-\frac{3l}{2}
\end{equation}

It is to be noted that the pseudoquantities given in  Eqs.
(38)-(43) are actual observables from the combined measurements of
rotation curves and gravitational lensing. Equation (43) provides
a convenient parameter that gives a 'measure' of the equation of
state of the halo field, calculated from a combination of rotation
curves and lensing measurements. Figure 2 shows that the value of
$\omega$ is negative indicating repulsion.

\begin{figure}[ptb]
\begin{center}
\vspace{0.3cm}\includegraphics[width=0.4\textwidth]{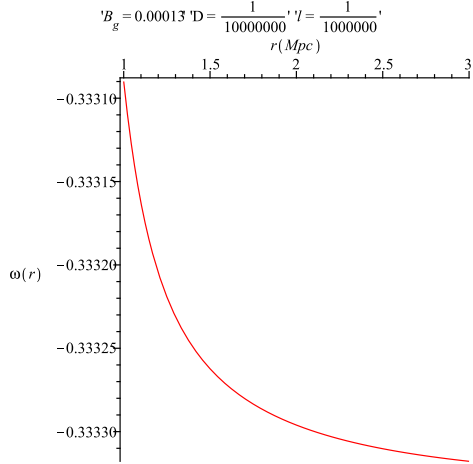}
\end{center}
\caption{The variation  $\omega(r)$ with respect to  r.}
\label{fig:4}
\end{figure}

\section{Conclusions}

We obtain the space time of the galactic core in the framework of
general relativity by assuming  the flat rotational curve
condition  and considering the matter content in the galactic core
region as strange quark matter. The spacetime metric is neither
asymtotically flat nor a spacetime due to a centrally symmetric
black hole. We also obtain the energy density, radial and
tangential pressures of the core strange quark matter. We see that
the NEC is satisfies. However, we note that the SEC is violated.
Therefore, it is obvious that quark matter is not purely real like
and it behaves like dark matter. Significantly we shown that Bag
Constant takes an important role to stabilize the circular orbit
of the test particles. We also give a limit of the Bag Constant
for the existence of quark matter in the galactic halo region.

\section*{Acknowledgments} We are thankful to Dr. G.C.Shit for helpful discussion. MK and FR  gratefully acknowledge support from the
Inter-University Centre for Astronomy and Astrophysics(IUCAA),
Pune, India under Visiting  Associateship programme under which a
part of this work was carried out.SMH is also thankful to IUCAA
for giving him an opportunity to visit IUCAA where a part of this
work was carried out. FR is personally thankful to PURSE,DST and
UGC for providing financial support.

 \end{document}